\documentclass[pra]{revtex4}
\usepackage{amsmath, amssymb, amsthm, amsfonts, latexsym, graphicx}
\usepackage{float}
\usepackage[caption=false]{subfig}
\usepackage{xcolor}
\usepackage{verbatim}
\usepackage{upgreek}
\usepackage{textgreek}
\input{Qcircuit}
\newcommand{\ket}[1]{\left| #1\right\rangle}        % ket vector
\newcommand{\bra}[1]{\left\langle #1\right|}        % bra vector
\newcommand{\ii}{\mathbb{I}}		% the I with two vertical lines
\newcommand{\norm}[1]{\left\| #1\right\|}        % norm
\newcommand{\proj}[1]{\ket{#1}\!\!\bra{#1}}

\newcommand{\oldbeta}{\zeta}

\newcommand{\maH}{H_{\text{max}}}
\newcommand{\mot}{\dot{H}_{\text{max}}}

\allowdisplaybreaks
\begin{document}
\title{Simulating the dynamics of time-dependent Hamiltonians  with~a~truncated~Dyson~series}
\author{M\'{a}ria Kieferov\'{a}}
\affiliation{%
	Department of Physics and Astronomy,
    Macquarie University,
    Sydney, NSW 2109,
    Australia.}
\affiliation{%
	Department of Physics and Astronomy, University of Waterloo and Institute for Quantum Computing, Ontario N2L 3G1, Canada}    
    
\author{Artur Scherer}
\affiliation{%
	Department of Physics and Astronomy,
    Macquarie University,
    Sydney, NSW 2109,
    Australia.}
\author{Dominic Berry }
\affiliation{%
	Department of Physics and Astronomy,
    Macquarie University,
    Sydney, NSW 2109,
    Australia.}

\begin{abstract}
We provide a general method for efficiently simulating time-dependent Hamiltonian dynamics
on a circuit-model based quantum computer. Our approach is based on approximating the truncated Dyson series
of the evolution operator, extending the earlier proposal by Berry \textit{et al.}~[Phys.~Rev.~Lett.\ \textbf{114}, 090502 (2015)] to evolution generated by explicitly time-dependent Hamiltonians.
Two alternative strategies are proposed to implement time ordering while 
exploiting the superposition principle for sampling the Hamiltonian at different times.
The resource cost of our simulation algorithm retains the optimal logarithmic dependence on the inverse of the desired precision.
\end{abstract}
\date{\today}
\maketitle

\section{Introduction}
Simulation of physical systems is envisioned to be one of the main applications for quantum computers~\cite{preskill2018}. Effective modeling of the dynamics and its generated time evolution is crucial to a deeper understanding of many-body systems, spin models, and quantum chemistry~\cite{abrams1997}, and 
may thus have significant implications for many areas of chemistry and materials sciences. 
Simulation of the intrinsic Hamiltonian evolution of quantum systems was the first potential use of quantum computers suggested by Feynman~\cite{Feynman1982} in 1982. Quantum simulation algorithms can model the evolution of a physical system with a complexity logarithmic in the dimension of the Hilbert space~\cite{georgescu2014} (i.e.\ polynomial in the number of particles), unlike classical algorithms whose complexity is typically polynomial in the dimension, making  simulations for practically interesting systems intractable for classical computers.

The first quantum simulation algorithm was proposed by Lloyd~\cite{Lloyd1073} in 1996.
There have been numerous advances since then providing improved performance~\cite{wiebe2010higher,wiebe2011simulating,childs2010relationship,PhysRevLett.106.170501,BerryQIC12,BerrySTOC14,BerryFOCS15,berry2015simulating,LowPRL2016,Low2016}.
One advance was to provide complexity that scales logarithmically in the error, and is nearly optimal in all other parameters~\cite{berry2015simulating}. 
Further improvements were provided by quantum signal processing methodology~\cite{LowPRX2016,LowPRL2016} as well as qubitization \cite{Low2016}, which
achieve optimal query complexity.

An important case is that of time-dependent Hamiltonians.
Efficient simulation of time-dependent Hamiltonians would allow us to devise better quantum control schemes~\cite{pang2017} and describe transition states of chemical reactions~\cite{ch_reactions}. Furthermore, simulation of dynamics generated by time-dependent Hamiltonians is a key component for implementing adiabatic algorithms~\cite{farhi2001quantum} and the quantum approximate optimization algorithm~\cite{farhi2014} in a gate-based quantum circuit architecture. 

The most recent advances in quantum simulation algorithms are for time-independent  Hamiltonians.
Techniques for simulating time-dependent Hamiltonians based on 
the Lie-Trotter-Suzuki decomposition were developed in~\cite{wiebe2011simulating,PhysRevLett.106.170501}, but the complexity scales polynomially with error.
More recent advances providing complexity logarithmic in the error \cite{BerrySTOC14,berry2015simulating} mention that their techniques can be generalized to time-dependent scenarios, but do not analyze this case.
The most recent algorithms \cite{LowPRL2016,Low2016} are not directly applicable to the time-dependent case.
Here we present an explicit algorithm for simulating time-dependent Hamiltonians with complexity logarithmic in the allowable error, matching the complexity of the algorithms for the time-independent case in \cite{BerrySTOC14,berry2015simulating}, though not those in \cite{LowPRL2016,Low2016}.

Our approach is based on a truncated Dyson series, similar to Ref.~\cite{berry2015simulating}.
Whereas Ref.~\cite{berry2015simulating} used a Taylor series, we here use a Dyson series to take account of time dependence of the Hamiltonians.
Our approach is also conceptually similar to Ref.~\cite{PhysRevLett.106.170501}, which showed that classically choosing random times could eliminate dependence on the derivatives of the Hamiltonian.
Our technique uses a quantum superposition of times, which achieves similar dependence on the derivatives (the complexity scales only as the log of the derivative of the Hamiltonian), but improves upon \cite{PhysRevLett.106.170501} by providing logarithmic scaling in the error.
We take the Hamiltonian to be given by an oracle for the entries in a sparse matrix, similar to Ref.~\cite{BerrySTOC14}.
That is so the Hamiltonian can be expressed as a linear combination of unitary terms.
It would also be possible to take the Hamiltonian to already be given as a linear combination of unitaries as in Ref.~\cite{berry2015simulating}.

This paper is organized as follows. In Sec.~\ref{sec:Results} we first state our main result.  
In Sec.~\ref{sec:background} we introduce our framework of definitions and assumptions.
Section~\ref{sec:Algorithm-Overview} then provides an overview of our simulation algorithm.
The main technical difficulty is in ordering the time register, which is presented in Sec.~\ref{sec:aux-reg-states}.
We give two alternative methods for implementing time-ordering using an additional control register for selecting the time variable values in the correct order.
Our first approach given in Sec.~\ref{sec:compressed-encoding} uses a compression technique to encode ordered sequences of times with minimal overhead.
Our second method is based on the quantum sort technique and presented in Sec.~\ref{sec:quantum-sort}. 
In Sec.~\ref{sec:state-preparation-part3} we show that oblivious amplitude amplification can be achieved with both approaches.
We derive the overall query and gate complexity in Sec.~\ref{sec:ComplexityReq}. 
We conclude with a brief summary and discussion of our results in Sec.~\ref{sec:Conclusion}. 
  
 \section{Main Result}
 \label{sec:Results}
We consider a time-dependent Hamiltonian $H(t)$ where the positions and values of nonzero entries are given by an oracle.
The Hamiltonian is $d$-sparse on the time interval $[t_0,t_0+T]$, in the sense that in any row or column there are at most $d$ entries that can be nonzero for any $t\in[t_0,t_0+T]$.
Moreover, we take $\epsilon$ to be the maximum allowable error of the simulation and $n$ to be the number of qubits for the quantum system on which $H(t)$ acts.
We define
\begin{equation}
\maH := \text{max}_{t\in[t_0,t_0+T]} \left|\left| H(t) \right|\right|_{\text{max}}, \qquad \mot := \text{max}_{t\in 
\left[t_0,t_0+T\right]} \norm{dH(t)/dt},
\end{equation}
where $\| \cdot \|$ indicates the spectral norm.
The evolution of a quantum system under the Hamiltonian $H(t)$ can be simulated for time $T$ and within error $\epsilon$ using a number of oracle queries scaling as
\begin{equation}
\mathcal{O}\left(d^2\maH T \frac{\log(d\maH T/\epsilon)}{\log\log(d\maH T/\epsilon)}\right),
\end{equation}
and a number of additional elementary gates scaling as
\begin{equation}
\mathcal{O}\left(d^2\maH T \frac{\log(d\maH T/\epsilon)}{\log\log(d\maH T/\epsilon)}\left[ \log\left(\frac{d \maH T}{\epsilon}\right) + \log\left( \frac{\mot T}{\epsilon  d \maH}\right) + n \right]\right).
\end{equation}
  
\section{Background and Framework}
\label{sec:background}

The unitary operator corresponding to time evolution 
under an explicitly time-dependent Hamiltonian $H(t)$ for a time period $T$ is the time-ordered exponential 
\begin{equation}
U(t_0, t_0 + T) = \mathcal{T} \exp{\left(-i\int_{t_0}^{t_0 + T} d\tau\, H(\tau)\right)}, \label{time-ordering}
\end{equation}
where $\mathcal{T}$ is the time-ordering operator and we 
use natural units ($\hbar=1$). 
Equation \eqref{time-ordering} can be understood as a limit
\begin{equation}
U(t_0, t_0 + T) =  \lim_{M \rightarrow \infty} \mathcal{T}  \prod_{n=0}^{M-1} \exp{\left\{\frac{-iT}{M}  H\!\left(t_0 + 
\frac{nT}{M}\right)\right\}},
\end{equation}
where $\mathcal{T}$ indicates the strict order of the terms in the product.

Access to the Hamiltonian is provided through the oracles
\begin{align}
O_{\text{loc}}\ket{j,s}&= \ket{j,\nu(j,s)}\,,\\
O_{\text{val}}\ket{t,i,j,z}&= \ket{t,i,j,z\oplus H_{ij}(t)}\,.
\end{align}
Here, $\nu(j,s)$ gives the position of the $s$'th element in row $j$ of $H(t)$ that may be nonzero at any time. Note that the oracle ${O}_{\text{loc}}$ does not depend on time. Oracle $O_{\text{val}}$  yields the values of non-zero elements at each instant of time. 
Furthermore, we say that a time-dependent Hamiltonian $H(t)$ is \emph{$d$-sparse on the interval} if the number of entries in any row or column that may be nonzero at any time throughout the interval is at most $d$.
This definition is distinct from the maximum sparseness of the instantaneous Hamiltonians $H(t)$, because some entries may go to zero while others become nonzero.
(This definition of sparseness upper bounds the sparseness of instantaneous Hamiltonians.)

Our algorithm builds on the unitary decomposition of a Hermitian matrix into equal-sized $1$-sparse self-inverse parts introduced in Lemma 4.3 in Ref.~\cite{BerrySTOC14}.
The Hamiltonian $H(t)$ can be decomposed using the technique of Ref.~\cite{BerrySTOC14} for any individual time, giving
\begin{equation} 
H(t) =\gamma \sum_{\ell=0}^{L-1} H_\ell(t),\label{unitary_decomposition2} 
\end{equation}
where the $H_\ell(t)$ are $1$-sparse, unitary and Hermitian.
The matrix entries in $H_\ell(t)$ can be determined using $\mathcal{O}(1)$ queries according to Lemma 4.4 of Ref.~\cite{BerrySTOC14}.
The single coefficient $\gamma$ is time-independent, 
and so is the sum of coefficients $\lambda := L\gamma$ in the decomposition.
To give a contribution to the error that is $\mathcal{O}(\epsilon)$, the value of $\gamma$ should be chosen as (see 
Eq.~(24) in \cite{BerrySTOC14} and the accompanying discussion)
\begin{equation}\label{gamchoice}
\gamma \in \Theta(\epsilon/(d^3 T)).
\end{equation}
The unitary decomposition \eqref{unitary_decomposition2} 
can be computed 
with a number of Hamiltonians scaling as
\begin{equation}\label{Lsca}
L\in\mathcal{O}\left(d^2 \maH/\gamma\right).
\end{equation}

As is common, we will quantify the resource requirements of the algorithm in terms
of two complexity measures: the `{\em query complexity}' and the `{\em gate  complexity}'. By the former measure we mean the number of queries to the oracles introduced above, i.e.\ the number of uses of the unitary procedures efficiently implementing these
oracles, thereby dismissing all details of their implementation and treating them as black boxes. By the latter measure we mean the total number of elementary 1- and 2-qubit gate operations.

\section{Algorithm Overview}
\label{sec:Algorithm-Overview}
Suppose we wish to simulate the action of $U$ in Eq.~\eqref{time-ordering} 
within error $\epsilon$. First, the total simulation time $T$ is divided into $r$ time segments of length $T/r$. Without loss of generality, we analyze the evolution induced within the first segment and set $t_0=0$.
The simulation of the evolution within the following segments is accomplished in the same way. 
The overall complexity of the algorithm is then given by the number
of segments $r$ times the cost of simulation for each segment. To upper bound 
the overall simulation error by $\epsilon$, 
we require the error of simulation for each segment to be at most $\epsilon/r$. 

We need a set of qubits to encode the time over the entire time interval $[0,T]$.
It is convenient to take $r$ to be a power of two, so there will be one set of qubits for the time that encodes the segment number, and another set of qubits that gives the time within the segment.
The qubits encoding the segment number are only changed by incrementing by $1$ between segments, which gives complexity $\mathcal{O}(r\log r)$.

Second, we approximate the 
evolution within the first segment by a Dyson series up to the $K$-th order: 
\begin{equation}
U(0,T/r)\approx\sum_{k=0}^K \frac{(-i)^k}{k!} \mathcal{T}\int_0^{T/r} d{\bf t}\, H(t_k) \dots H(t_1)\;, 
\label{integral}
\end{equation}
where, for each $k$-th term in the Dyson series, $\mathcal{T} \int_0^{T/r}  d{\bf t}\,(\cdot)\,$ represents integration over a $k$-tuple of time variables $(t_1,\dots, t_k)$ while keeping the times ordered: $t_1\leq t_2\leq\dots\leq t_k$. It is straightforward to show that the error of this approximation is 
$\mathcal{O}\left(\frac{\maH^{K+1}}{(K+1)!} 
\left(\frac{T}{r}\right)^{K+1} \right)$. 

Next, we discretize the integral over each time variable and approximate it by a sum 
with $M$ terms. The time-dependent Hamiltonian is thereby approximated by its values at $M$ 
uniformly separated times $\frac{jT}{rM}$ identified by an integer $j\in\{0,\dots ,M-1\}$.
\begin{comment}

\end{comment}
Replacing all integrals by sums in expression \eqref{integral}  yields the following approximation of the time-evolution operator within the first segment:
\begin{equation}
\widetilde{U} =\sum_{k=0}^K \frac{(-iT/r)^k}{M^k k!} \sum_{j_1, \dots, j_k =0} ^{M-1} \mathcal{T} H(t_{j_k}) \dots  
H(t_{j_1})\label{taylor_approx}.
\end{equation}
Replacing the integrals by Riemann sums contributes an additional error 
upper bounded by $\mathcal{O} 
\left(  \frac{ (T/r)^2 \dot{H}_{\text{max}}}{M} \right)$ (see Sec.~\ref{sec:ComplexityReq} for more details).
The overall 
error of the obtained approximation is thus 
\begin{equation}
\|\widetilde{U}-U(0,T/r)\|\in\mathcal{O} 
\left( \frac{{(\maH T/r)}^{K+1}}{(K+1)!} + \frac{ (T/r)^2 \dot{H}_{\text{max}}}{M} \right)\,.
\end{equation}
Provided that $r\ge \maH T$, the overall error can be bounded by $\epsilon/r$ if we choose
\begin{eqnarray}
&&K\in \Theta\left( \frac{\log{(r/\epsilon)}}{\log\log{(r/\epsilon)}} \right)\quad\text{and}\quad M\in \Theta\left( 
\frac{T^2  \dot{H}_{\text{max}}}{\epsilon r } \right)
\,.
\label{eq:scaling_KM}
\end{eqnarray}
These expressions also imply that $M$ should be exponentially larger than $K$ in terms of the error $\epsilon / r$. 

The main difference from time-independent Hamiltonians is that for time-dependent Hamiltonians we need to implement the evolution generated 
by $H(t)$ for different times in the correct order. We achieve this by introducing an additional multi-qubit control register called `\texttt{time}'. This ancilla register is 
prepared in a certain superposition state 
(depending on which approach we take). It is used to sample the Hamiltonian at different times in superposition 
in a way that respects time ordering.

Substituting the unitary decomposition~\eqref{unitary_decomposition2} into~\eqref{taylor_approx}, the approximation of the time-evolution operator within the first segment takes the form $\tilde{U}= \sum_{\boldsymbol{j}\in J} \beta_{\boldsymbol{j}} V_{\boldsymbol{j}}$, where $\boldsymbol{j}$ is a multi-index and the coefficients  $\beta_{\boldsymbol{j}}$ comprise information about both the time discretization and the unitary decomposition 
weightings as well as the order $k$ within the Taylor series. Explicitly, we define  
\begin{align}\label{betas}
\beta_{\left(k,\ell_1,\dots,\ell_k, j_1,\dots,j_k\right)}&:=\frac{\left(\gamma\, T/r\right)^k}{M^k k_1!k_2!\ldots k_\sigma!}\uptheta_k
\left({j_1},\dots,{j_k}\right),
\\
V_{\left(k,\ell_1,\dots,\ell_k,j_1,\dots,j_k\right)}&:=(-i)^k 
\;H_{\ell_k}(t_{j_k})\dots H_{\ell_1}(t_{j_1})\,,
\end{align} 
where $\uptheta_k\left({j_1},\dots,{j_k}\right)=1$ if $j_1\le j_2\le\ldots\le j_k$, and zero otherwise.
The quantity $\sigma$ is the number of distinct values of $j$, and $k_1, k_2, \ldots, k_\sigma$ are the number of repetitions for each distinct value of $j$.
That is, we have the indices $j$ for the times sorted in ascending order, and we have multiplied by a factor of $k!/(k_1!k_2!\ldots k_\sigma!)$ to take account of the number of unordered sets of indices which give the same ordered set of indices.
The multi-index set $J$ is defined as 
\begin{equation}
J:=\left\{\left(k,\ell_1,\dots,\ell_k,j_1,\dots,j_k\right)\,:\,k\in \{0,\dots,K\}, \;\ell_1,\dots,\ell_k\in \{0,\dots,L-1\},\;j_1,\dots, j_k\in\{0,\dots,M-1\}\right\}.
\label{eq:Def-Index-J}
\end{equation}

We use a standard technique to implement linear combinations of unitaries (referred to as `LCU technique') involving the use of an ancillary register to encode the 
coefficients $\beta_{\boldsymbol{j}}$ \cite{BerrySTOC14}.
In the next section, we present two approaches to  implementation of the (multi-qubit) ancilla state preparation  
\begin{equation} B\ket{0}_a=\frac{1}{\sqrt{s}}\sum_{\boldsymbol{j}\in J}\sqrt{\beta_{\boldsymbol{j}}}\ket{\boldsymbol{j}}_a
\label{eq:state-preparation}
\end{equation}
as part of the LCU approach, where $s:=\sum_{\boldsymbol{j}\in J} \beta_{\boldsymbol{j}}$.
For the LCU technique we introduce the operator $\textsc{Select}(V):=\sum_{\boldsymbol{j}}\proj{\boldsymbol{j}}_a\otimes V_{\boldsymbol{j}}$ acting as 
\begin{align}
\textsc{Select}(V)\ket{\boldsymbol{j}}_a\ket{\psi}_s=\ket{\boldsymbol{j}}_aV_{\boldsymbol{j}}\ket{\psi}_s
\end{align}
on the joint ancilla and system states. This operation implements a term from the decomposition of $\tilde{U}$ 
{\em selected} by the ancilla state $\ket{\boldsymbol{j}}_a$ 
with weight $\beta_{\boldsymbol{j}}$.
Following the method in 
Ref.~\cite{berry2015simulating}, we also define  
\begin{align}
R&:=\ii_{as} - 2\left(\ket{0}\!\bra{0}_a\otimes\ii_s \right), \\
W&:=\left(B^{\dagger} \otimes \ii_s\right) \textsc{Select}(V)  \left(B \otimes \ii_s\right).
\label{eq:define-W}
\end{align}
If $\widetilde{U}$ were unitary and $s\le 2$, a single step of oblivious amplitude amplification could be used to implement $\widetilde{U}$.
When $\widetilde{U}$ is only approximately unitary, with $\|\widetilde{U}-U(0,T/r)\|\in\mathcal{O}(\epsilon/r)$, a single step of oblivious amplitude amplification yields \cite{berry2015simulating}
\begin{align}
-WRW^{\dagger}RW\ket{0}_a\ket{\psi}_s=\ket{0}_a\tilde{U}\ket{\psi}_s + \mathcal{O}\left(\epsilon/r\right),
\label{eq:OAA-transformation}
\end{align}
which is the approximate transformation we aim to implement for each time segment.

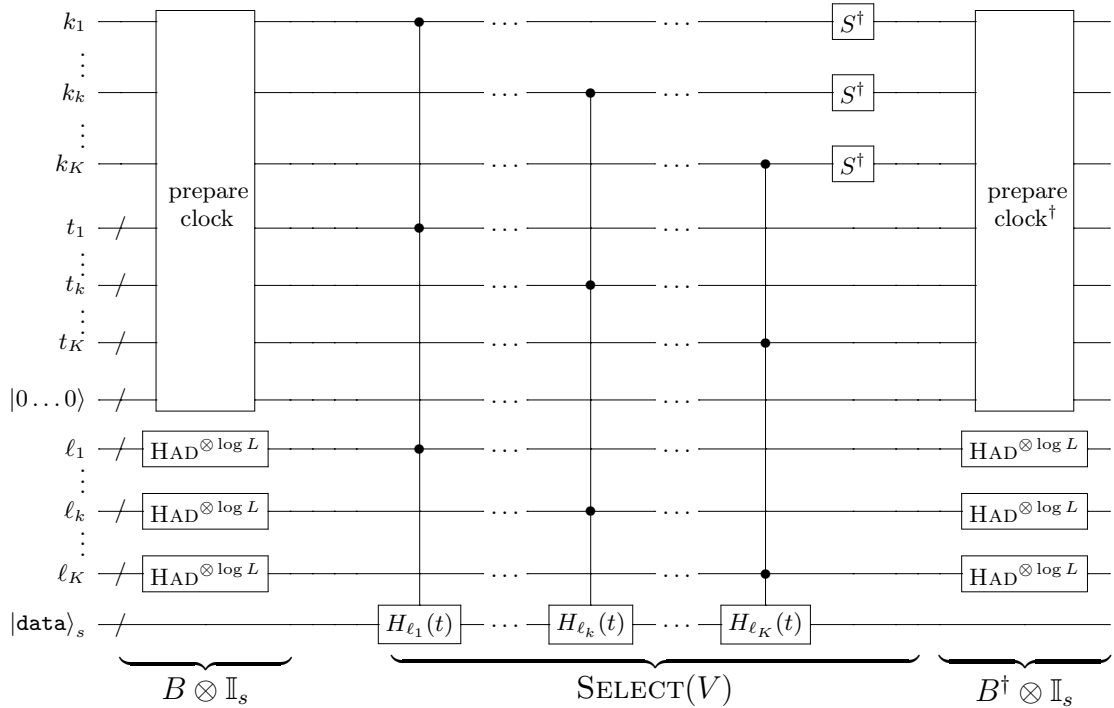
\begin{figure}[bth]{
\centering
%{\scriptsize
$$
\hspace{10mm}
\Qcircuit @C=0.9em @R=0.273em { 
\lstick{k_1}            &\qw&\multigate{12}{\begin{minipage}{3.2em}\!\! prepare\newline \ \ clock\end{minipage}\!} 
&\qw&\qw&\qw&\qw&\ctrl{6}  &\qw &\dots &           &\qw &\qw &\dots &     &\qw &\gate{S^\dagger}  &\qw         &\qw &\qw  &\multigate{12}{\begin{minipage}{3.2em}\!\! prepare\newline \ \ clock$^{\dagger}$\end{minipage}\!}   &\qw\\
 \lstick{\vdots}    & & \pureghost{\begin{minipage}{3.2em}\!prepare\newline clock\end{minipage}\!}    &                        &                        &     &        & &                        &     &        & &                  &          &  \\
\lstick{k_k}           &\qw&\ghost{\begin{minipage}{3.2em}\!prepare\newline clock\end{minipage}\!}
&\qw&\qw&\qw&\qw&\qw                &\qw &\dots &           &\ctrl{6} &\qw &\dots &     &\qw  &\gate{S^\dagger}  &\qw         &\qw  & \qw&\ghost{\begin{minipage}{3.2em}\!prepare\newline clock\end{minipage}\!}    &\qw \\
 \lstick{\vdots}    & & \pureghost{\begin{minipage}{3.2em}\!prepare\newline clock\end{minipage}\!}    &                        &                        &     &        & &                        &     &        & &                  &          &  \\
\lstick{k_K}   &\qw&\ghost{\begin{minipage}{3.2em}prepare\newline clock\end{minipage}\!}
&\qw&\qw&\qw&\qw&\qw                      &\qw &\dots &           &\qw &\qw &\dots &     &\ctrl{6}  &\gate{S^\dagger}    &\qw    &\qw   &\qw&\ghost{\begin{minipage}{3.2em}\!prepare\newline clock\end{minipage}\!}     &\qw              \\
 \lstick{}    &  & \pureghost{\begin{minipage}{3.2em}\!prepare\newline clock\end{minipage}\!}    &                        &                        &     &        & &                        &     &        & &                  &          &  \\
\lstick{t_1}           &{/}\qw&\ghost{\begin{minipage}{3.2em} prepare\newline  clock \end{minipage}\!}
&\qw&\qw&\qw&\qw&\ctrl{9}    &\qw &\dots &           &\qw &\qw &\dots &     &\qw  &\qw   &\qw         &\qw  &\qw  &\ghost{\begin{minipage}{3.2em}\!prepare\newline clock\end{minipage}\!}     &\qw\\
 \lstick{\vdots}    & & \pureghost{\begin{minipage}{3.2em}\!prepare\newline clock\end{minipage}\!}    &                        &                        &    &        & &                       &     &        & &                  &          &  \\
\lstick{t_k}           &{/}\qw&\ghost{\begin{minipage}{3.2em} prepare\newline  clock \end{minipage}\!}
&\qw&\qw&\qw&\qw&\qw                      &\qw &\dots &           &\ctrl{11} &\qw &\dots &     &\qw  &\qw    &\qw     &\qw  &\qw  &\ghost{\begin{minipage}{3.2em}\!prepare\newline clock\end{minipage}\!}   &\qw \\
 \lstick{\vdots}    & & \pureghost{\begin{minipage}{3.2em}\!prepare\newline clock\end{minipage}\!}    &                        &                        &     &        & &                        &     &        & &                  &          &  \\
\lstick{t_K}           &{/}\qw&\ghost{\begin{minipage}{3.2em} prepare\newline  clock \end{minipage}\!}
&\qw&\qw&\qw&\qw&\qw                       &\qw &\dots &           &\qw &\qw &\dots &     &\ctrl{13}  &\qw    &\qw    &\qw  &\qw  &\ghost{\begin{minipage}{3.2em}\!prepare\newline clock\end{minipage}\!}    &\qw \\
 \lstick{}    & & \pureghost{\begin{minipage}{3.2em}\!prepare\newline clock\end{minipage}\!}    &                        &                        &     &        & &                        &     &        & &                  &          &  \\
\lstick{\ket{0\dots 0}}&{/}\qw&\ghost{\begin{minipage}{3.2em} prepare\newline  clock \end{minipage}\!}
&\qw&\qw&\qw&\qw&\qw                            &\qw &\dots &           &\qw &\qw &\dots &     &\qw  &\qw    &\qw    &\qw  &\qw  &\ghost{\begin{minipage}{3.2em}\!prepare\newline clock\end{minipage}\!}    &\qw\\
\lstick{} & &    &                        &                        &     &        & &                        &     &        & &                  &          &  \\
\lstick{} & &    &                        &                        &     &        & &                        &     &        & &                  &          &  \\
\lstick{\ell_1}           &{/}\qw&\gate{\textsc{Had}^{\otimes \log L}}  
&\qw &\qw&  \qw             &\qw&\ctrl{10}&\qw &\dots &   &\qw &\qw &\dots &     &\qw  &\qw  &\qw         &\qw  &\qw  &\gate{\textsc{Had}^{\otimes  \log L}}        &\qw\\
 \lstick{\vdots}    & &    &                        &                        &     &        & &                        &     &        & &                  &          &  \\
 \lstick{}           &&   
&&&            &&      & & &           & & & &     &  &  &      &  &  &   &  \\
 \lstick{}           &&   
&&&            &&      & & &           & & & &     &  &  &      &  &  &   &  \\
\lstick{\ell_k}           &{/}\qw&\gate{\textsc{Had}^{\otimes \log L}}   
&\qw &\qw&    \qw         &\qw&\qw      &\qw &\dots &           &\ctrl{6} &\qw &\dots &     &\qw  &\qw  &\qw      &\qw  &\qw  &\gate{\textsc{Had}^{\otimes \log L}}     & \qw \\
 \lstick{\vdots}           &&   
&&&            &&      & & &           & & & &     &  &  &      &  &  &   &  \\
 \lstick{}           &&   
&&&            &&      & & &           & & & &     &  &  &      &  &  &   &  \\
 \lstick{}           &&   
&&&            &&      & & &           & & & &     &  &  &      &  &  &   &  \\
\lstick{\ell_K}           &{/}\qw&\gate{\textsc{Had}^{\otimes \log L}}    
&\qw & \qw &\qw&\qw&\qw       &\qw &\dots &           &\qw &\qw &\dots &     &\ctrl{2}      
&\qw &\qw  &\qw  &\qw  &\gate{\textsc{Had}^{\otimes \log L}}    &\qw      \\
 \lstick{}           &&   
&&&            &&      & & &           & & & &     &  &  &      &  &  &   &  \\
\lstick{\ket{\texttt{data}}_s}    &{/}\qw&\qw   
&\qw&\qw&\qw&\qw                                &\gate{H_{\ell_1}(t) }&\qw &\dots & &\gate{H_{\ell_k}(t) }&\qw  &\dots &   &\gate{H_{\ell_K}(t) }   &\qw 
&\qw    &\qw      &\qw      &\qw  &\qw  \\
 \lstick{}           &&   
&&&            &&      & & &           & & & &     &  &  &      &  &  &   &  \\
 \lstick{}           &&   
&&&            &&      & & &           & & & &     &  &  &      &  &  &   &  \\
 \lstick{}           &&   
&&&            &&      & & &           & & & &     &  &  &      &  &  &   &  \\
 \lstick{}           &&   
&&&            &&      & & &           & & & &     &  &  &      &  &  &   &  \\
 \lstick{}           &&  \relax\underbrace{\quad\quad\quad\quad\quad\quad\quad}_{\text{\large $ B\otimes \ii_s$}}
&&&&&&&& &&  \relax\underbrace{\;\;\quad\quad\quad\quad\quad\quad\quad\quad\quad\quad\quad\quad\quad\quad\quad\quad\quad\quad\quad\quad\quad}_{\text{\large $\textsc{Select}(V)$}}         &&      & & &            & &     &\relax\underbrace{\quad\quad\quad\quad\quad\quad\quad}_{\text{\large $ B^\dagger \otimes \ii_s$}}  &  &  &     &  &  &   &  \\
}
$$
%}
}
	\caption{%
    Quantum circuit implementing the unitary transformation $W$ as defined in Eq.~\eqref{eq:define-W}.  
The boxed subroutine `prepare clock' 
is implemented either by the `compressed encoding' 
approach outlined in Sec.~\ref{compression}, or by using a quantum sorting network described in  Sec.~\ref{sec:quantum-sort}.
The phase gate 
$S^\dagger:=\proj{0}+(-i)\proj{1}$ is used to implement 
the factor $(-i)^k$ as part of the $k$-th order term of the Dyson series. \label{fig:QCircuit-W}}
\end{figure}

The implementation of the unitary transformation $W$ by a quantum circuit is illustrated in Fig.~\ref{fig:QCircuit-W}. It 
generalizes the LCU technique to time-dependent decompositions. In addition to the \texttt{data} register holding the data to be processed by Hamiltonian evolution, three auxiliary control registers are employed.   
The `\texttt{k} register' consisting of $K$ single-qubit wires is used to hold the $k$ value corresponding to the order to be taken within the truncated  Dyson series.  The \texttt{time} register consisting of $K$ subregisters each of size $\lceil \log M\rceil$ is prepared in a special superposition state  `\texttt{clock}' that also involves the $\texttt{k}$ 
register and which is used to sample the Hamiltonian at different instances of time in superposition such that time-ordering is accounted for. Finally, 
the `$\texttt{l}$ register' consisting of $K$ subregisters each of size $\lceil \log L\rceil$ is used to prepare the ancillary register states commonly needed  
to implement the LCU technique. Its task is to select 
which term out of the decomposition into unitaries is to be applied. 
It is convenient to take both $L$ and $M$ to be powers of two, so equal-weight superpositions can be produced with tensor products of Hadamards, for example
$\textsc{Had}^{\otimes  \log L}$ for the $\texttt{l}$ register.
Here and throughout the paper, all logarithms are taken to the base 2.

\section{Preparation of auxiliary register states}
\label{sec:aux-reg-states}
The algorithm relies on efficient implementation of the unitary 
transformation $B$ to prepare ancilla registers in the superposition states given in Eq.~\eqref{eq:state-preparation}. As noted above, 
the main difficulty of simulating time-dependent Hamiltonian dynamics 
is the implementation of time-ordering. This is achieved by a weighted superposition named  `\texttt{clock}'
prepared in an auxiliary 
control register named `\texttt{time}', 
which is introduced in addition to the ancilla registers used to implement the LCU technique in the time-independent case. 
Its key purpose and task is to control sampling the Hamiltonian for different instances of time in superposition in a way that respects the correct time order. We here present two 
alternative approaches to efficient preparation of such clock states in the \texttt{time} register. 

The first approach is based upon generating the intervals between successive times according to an exponential distribution, in a similar way as in Ref.~\cite{berry2014gate}.
These intervals are then summed to obtain the times.
Our second approach starts by creating a superposition over $k$ and then a superposition over all possible times $t_1, \dots , t_k$ for each $k$. The times are then reversibly sorted using quantum sorting networks \cite{cheng2006quantum,beals2013efficient} and used to control $\ell$ as in the previous approach. 

\subsection{Clock Preparation Using Compressed Rotation Encoding}\label{compression}
\label{sec:compressed-encoding}
To explain the first approach, it is convenient to consider a conceptually simple but inefficient representation of the time register.
An ordered sequence of times $t_1, \dots, t_k$ is encoded in a binary string $x\in\{0,1\}^M$ with Hamming weight $|x|=k$, where $0\leq k\leq K$.
That is, if $m_j$ is the $j$'th value of $m$ such that $x_m=1$, then $t_j={m_jT}/(rM)$.
This automatically forces the ordering $t_1<t_2<\dots<t_k$, omitting the terms when two or more Hamiltonians act at the same 
time. The binary string $x$ then would be encoded into $M$ qubits as $\ket{x}$. 

Now consider these $M$ qubits initialized to $\ket{0}^{\otimes M}$, then rotated by a small angle, to give
\begin{align}
\left( \frac{\ket{0}+ \sqrt{\oldbeta}\ket{1}}{\sqrt{1+\oldbeta}} \right)^{\otimes M}&= \left(1+\oldbeta\right)^{-M/2} \sum_{x} \oldbeta^{\left|x\right|/2} \ket{x}\nonumber \\
&= \left(1+\oldbeta\right)^{-M/2} \sum_{x, \left|x\right|\leq K} \oldbeta^{\left|x \right|/2} \ket{x} + \left(1+\oldbeta\right)^{-M/2} \sum_{x, \left|x\right|>K} \oldbeta^{\left|x\right|/2} \ket{x}\nonumber \\
&= \sqrt{1-\mu^2} \ket{\texttt{time}} + \mu \ket{\nu} \label{rotation}
\end{align}
where $\oldbeta := \lambda T/(r M)$, and
\begin{equation}
\ket{\texttt{time}} =\frac{1}{\sqrt S} 
\sum_{\left|x\right|\leq K} 
\oldbeta^{\left|x\right|/2} \ket{x}, \label{time}
\end{equation}
with $S:=\sum_{\left|x\right|\leq K} 
\oldbeta^{\left|x\right|}$. 
The states $\ket{\texttt{time}}$ and $\ket{\nu}$ are normalized and orthogonal. 
The state $\ket{\texttt{time}}$ includes up to $K$ times, and is analogous to the terms up to order $K$ in the Dyson series. 
The state $\ket{\nu}$ is analogous to the higher-order terms omitted in the Dyson series.
The amplitude $\mu$ satisfies
\begin{equation}
 \mu^2 =\mathcal{O}\left(\frac {( {\lambda T}/{r})^{K+1}}{(K+1)!}\right).
\end{equation}
We choose $K$ as in Eq.~\eqref{eq:scaling_KM}, and we will choose $r>\lambda T$, in which case we obtain $\mu^2=\mathcal{O}(\epsilon/r)$.

Since $\ket{\texttt{time}}$ includes only Hamming weights up to $K$, it includes strings that are highly compressible.
In order to compress the string $x$, the lengths of the strings of zeros between successive ones are stored.
That is, a string $x=0^{s_1}10^{s_2}10^{s_3}\dots 0^{s_k}10^{t}$ is represented by the integers $s_1s_2\dots s_k$.
There is always a total of $K$ entries, regardless of the Hamming weight.
Instead of explicitly recording the Hamming weight, the state preparation procedure of Ref.~\cite{berry2014gate} indicates that there are no further ones using the entry $M$.
For example, for $K=3$, the state $\ket{001000001000}$ would be encoded as $\ket{2,5,12}$.
This encoding is denoted $C_M^K$, so
\begin{equation}
C_M^K\ket{x} = \ket{s_1, \dots, s_k, M,\dots, M}.
\end{equation}
According to Theorem 2 of \cite{berry2014gate}, the state
\begin{equation}
\ket{\Xi_M^K}=\sqrt{1-\mu^2} C_M^K\ket{\texttt{time}} + \mu \ket{\nu'}
\end{equation}
can be prepared within trace distance $\mathcal{O}(\delta)$ using a number of gates $\mathcal{O}(K[\log M+\log\log(1/\delta)])$.
Taking $\delta=\epsilon/r$, that means we can prepare the state $\ket{\texttt{time}}$ within trace distance $\mathcal{O}(\epsilon/r)$ using a number of gates $\mathcal{O}(K[\log M+\log\log(r/\epsilon)])$.

This preparation does not give us the state encoded in quite the way we want, because we want an additional register encoding $k$, and the absolute times, rather than the differences between the times.
First we can count the number of times $M$ appears to determine $k$, which has complexity $\mathcal{O}(K \log{M})$.
Then we can increment registers $2$ to $k$ to give $\ket{s_1,s_2+1 \dots, s_k+1, M,\dots, M} \ket{k}$.
Then we can add register $1$ to register $2$, then register $2$ to register $3$, and so forth up to adding register $k-1$ to register $k$, to give $\ket{j_1, \dots, j_k, M,\dots, M} \ket{k}$ with $j_1=s_1$, $j_2= s_1+s_2+1$, up to $j_k=s_1+s_2+\dots+ s_k + k-1$, to give times $t_j$. This again has complexity $\mathcal{O}(K \log{M})$ gates, giving a total complexity $\mathcal{O}(K[\log M+\log\log(r/\epsilon)])$.

For completeness we briefly describe the approach used in Ref.~\cite{berry2014gate} to prepare the state $\ket{\Xi_M^K}$.
The idea is that one prepares a state denoted $\ket{\phi_q}$ (see Eq.~(13) of \cite{berry2014gate}) which gives an exponential distribution.
The state $\ket{\phi_q}$ is given by
\begin{equation}
\ket{\phi_q} = \sum_{s=0}^{q-1} \beta \alpha^s \ket{s} + \alpha^q \ket{q},
\end{equation}
where $\alpha:=1/\sqrt{1+\oldbeta}$ and $\beta:=\sqrt{\oldbeta}\alpha$ (not related to the $\alpha$ and $\beta$ otherwise used here).
The value of $q$ is chosen as $\log q = \Theta(\log M+ \log\log(1/\delta))$.
This state can be used for the position of the first one, then another $\ket{\phi_q}$ gives the spacing between the first one and the second one, and so forth.
Taking a tensor product of $K+1$ of the states $\ket{\phi_q}$ then gives something similar to $\ket{\Xi_M^K}$, except there is a difficulty with computational basis states giving positions for ones past $M$.
In Ref.~\cite{berry2014gate} the proof of Lemma 3 explains how to clean up these cases to give the state encoded in the form given in $\ket{\Xi_M^K}$.

\subsection{Clock Preparation Using a Quantum Sort}
\label{sec:quantum-sort}
In this section, we explain an alternative approach 
to implementing the preparation of the \texttt{clock} state, 
which establishes time ordering via a reversible sorting algorithm.  
This approach first creates a superposition over all Dyson series orders $k\le K$ with amplitudes  $\propto\left(\oldbeta^k/k!\right)^{1/2}$ (recall $\oldbeta=\lambda T/(rM)$). It then generates a superposition over all possible $k$-tuples of times 
$t_1, \dots , t_k$ for each possible value of $k$. 
To achieve time-ordered 
combinations, sorting is applied to each of the tuples in the 
superposition using a quantum sorting network algorithm.

The \texttt{clock} preparation requires $\log{M}$ qubits to encode each value $t_j$ (via the integer $j$). As $k\le K$, the \texttt{time} register thus consists of 
$K \log{M}$ ancilla qubits. The value of $k$ is encoded in unary into $K$ additional qubits constituting the `\texttt{k}-register'. Additionally, further $\mathcal{O}\left(K \text{poly}(\log{K}) \right)$ ancillae are required to reversibly perform a quantum sort.
The overall space complexity thus amounts to $\mathcal{O}\left(  K\log{M}+ K\text{poly}(\log{K}) \right)$.

To be more concrete, in the first step we create a superposition over the allowed values of $k$ in unary encoding by applying the following transformation to $K$ qubits initialized to $\ket{0}$: 
\begin{equation}
\textsc{Prepare}(k) \ket{0}^{\otimes K}: = \frac{1}{\sqrt{s}}\sum_{k=0}^K \sqrt{\frac{\oldbeta^k M^k}{k!}}\ket{1^k0^{K-k}},
\end{equation}
where the constant $s$ accounts for normalization.
This  transformation can be easily implemented using a series of $\mathcal{O}(K)$ controlled rotations, namely, by first rotating the first qubit followed by rotations applied to qubits $k=2$ to $K$ 
controlled by qubit $k-1$, respectively. 
In what follows, we denote 
the state $ \ket{1^k0^{K-k}}$ simply as $\ket{k}$. We use it to determine which of the times $t_j$ satisfy the condition $j\leq k$. 

Next we wish to create an equal superposition over the time indices $j$.
We take $M$ to be a power of $2$, so the preparation can be performed via the Hadamard transforms $\textsc{Had}^{\otimes \log{M}}$ on all $K$ 
subregisters of \texttt{time} (each of size $\log{M}$) 
to generate the superposition state
\begin{equation}
\frac{1}{\sqrt{s}}\sum_{k=0}^K  \sqrt{\frac{\oldbeta^k M^k}{M^K k!}} \ket{k} \sum_{j_1=0}^{M-1} \sum_{j_2=0}^{M-1}  \dots \sum_{j_K=0}^{M-1} \ket{j_1, j_2, \dots, 
j_K} \label{all_times}
\end{equation}
using $\mathcal{O}(K \log{M})$ gates. (The same complexity can be obtained if $M$ is not a power of $2$, though the circuit is more complicated.)
At this stage, we have created all possible $K$-tuples of times 
without accounting for time-ordering. Note that this superposition is over all possible $K$-tuples of times, not only $k\leq K$ of them.  
We have a factor of $M^k/M^K$, but for any $k$ in the superposition we ignore the times in subregisters ${k+1}$ to $K$. Hence the amplitude for $\ket{j_1, j_2, \dots, j_k}$ is $\propto\sqrt{\oldbeta^k/k!}$.

The next step is to sort the values stored in the \texttt{time} subregisters.
Common classical sorting algorithms are often inappropriate for quantum algorithms, because they involve actions on registers that depend on the values found in earlier steps of the algorithm.
Sorting techniques that are suitable to be adapted to quantum algorithms are {\em sorting networks}, because they involve actions on registers in a sequence that is independent of the values stored in the registers. 
Methods to adapt sorting networks to quantum algorithms are discussed in detail in Refs.~\cite{cheng2006quantum,beals2013efficient}.

A sorting network involves a sequence of comparators, where each comparator compares the values in two registers and swaps them conditional on the result of the comparison.
If used as part of a quantum algorithm, 
a sorting network must be made reversible. 
This is achieved by recording the result of the comparison in an ancilla qubit. 
To be more specific, first the comparison operation 
$\textsc{Compare}$ acts on two multi-qubit registers storing 
the values $q_1$ and $q_2$ and an ancilla 
initialized in state $\ket{0}$ as follows:
\begin{equation}
\textsc{Compare}\ket{q_1}\ket{ q_2}\ket{0} = \ket{q_1}\ket{q_2} \ket{\uptheta(q_1-q_2)},
\end{equation}
where $\uptheta$ is the Heaviside step function 
(here using the convention $\uptheta(0)=0$). In other words, $\textsc{Compare}$ flips an ancilla if and only if $q_1>q_2$. Such a comparison of two $\log M$-sized registers can be implemented with $\mathcal{O}(\log M)$ elementary gates \cite{antisym}. The overall action of a comparator module is completed by conditionally swapping 
the values of the two compared  multi-qubit registers. This is implemented by 
$\textsc{Swap}$ gates controlled by the ancilla state.

For each $\ket{k}$ in the superposition \eqref{all_times}, we only want to sort the values in the first $k$ subregisters of \texttt{time}. 
We could perform $K$ sorts for each value of $k$, or, alternatively, control the action of the comparators such that they perform the $\textsc{Swap}$ only for subregisters with positions up to value $k$.
A more efficient approach is to
sort all the registers regardless of the value of $k$, and also perform the same controlled-swaps on the registers encoding the value of $k$ in unary.
This means that the qubits encoding $k$ still indicate whether the corresponding time register includes a time we wish to use.
The controlled $H_\ell(t)$ operations are controlled on the individual qubits encoding $k$, and will still be performed correctly.

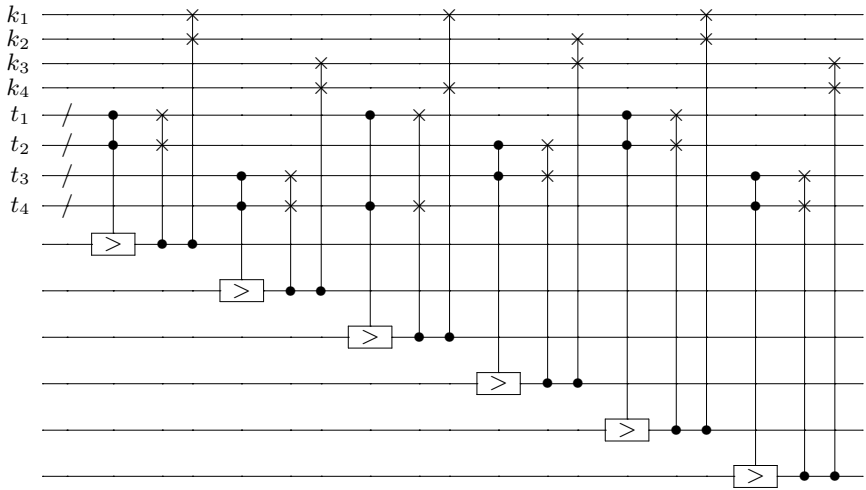
\begin{figure}[tbh]
$$
\Qcircuit @C=1em @R=1em {
\lstick{k_1} &\qw&\qw&\qw&\qswap&\qw&\qw&\qw           &\qw&\qw&\qswap&\qw&\qw&\qw                        &\qw&\qw&\qswap&\qw&\qw&\qw                        &\qw\\
\lstick{k_2} &\qw&\qw&\qw&\qswap&\qw&\qw&\qw           &\qw&\qw&\qw&\qw&\qw&\qswap                        &\qw&\qw&\qswap&\qw&\qw&\qw                        &\qw\\
\lstick{k_3} &\qw&\qw&\qw&\qw&\qw&\qw&\qswap           &\qw&\qw&\qw&\qw&\qw&\qswap                        &\qw&\qw&\qw&\qw&\qw&\qswap                        &\qw\\ 
\lstick{k_4} &\qw&\qw&\qw&\qw&\qw&\qw&\qswap           &\qw&\qw&\qswap&\qw&\qw&\qw                        &\qw&\qw&\qw&\qw&\qw&\qswap                        &\qw\\ 
\lstick{t_1} &{/} \qw&\ctrl{1}&\qswap&\qw&\qw&\qw&\qw  &\ctrl{3}&\qswap&\qw&\qw&\qw&\qw                   &\ctrl{1}&\qswap&\qw&\qw&\qw&\qw                   &\qw\\
\lstick{t_2} &{/} \qw&\ctrl{3}&\qswap&\qw&\qw&\qw&\qw  &\qw&\qw&\qw&\ctrl{1}&\qswap&\qw                   &\ctrl{7}&\qswap&\qw&\qw&\qw&\qw                   &\qw\\
\lstick{t_3} &{/} \qw&\qw&\qw&\qw&\ctrl{1}&\qswap&\qw  &\qw&\qw&\qw&\ctrl{5}&\qswap&\qw                   &\qw&\qw&\qw &\ctrl{1}&\qswap&\qw                  &\qw\\
\lstick{t_4} &{/} \qw&\qw&\qw&\qw&\ctrl{2}&\qswap&\qw  &\ctrl{3}&\qswap&\qw&\qw&\qw&\qw                   &\qw&\qw&\qw &\ctrl{6}&\qswap&\qw                  &\qw\\
&\qw&\multigate{0}{>}&\ctrl{-4}&\ctrl{-8}&\qw&\qw&\qw  &\qw&\qw&\qw&\qw&\qw&\qw                           &\qw&\qw&\qw&\qw&\qw&\qw                           &\qw\\
&\qw&\qw&\qw&\qw&\multigate{0}{>}&\ctrl{-3}&\ctrl{-7}  &\qw&\qw&\qw&\qw&\qw&\qw                           &\qw&\qw&\qw&\qw&\qw&\qw                           &\qw\\
&\qw&\qw&\qw&\qw&\qw&\qw&\qw                           &\multigate{0}{>}&\ctrl{-6}&\ctrl{-10}&\qw&\qw&\qw &\qw&\qw&\qw&\qw&\qw&\qw                           &\qw\\
&\qw&\qw&\qw&\qw&\qw&\qw&\qw                           &\qw&\qw&\qw&\multigate{0}{>}&\ctrl{-6}&\ctrl{-10} &\qw&\qw&\qw&\qw&\qw&\qw                           &\qw\\
&\qw&\qw&\qw&\qw&\qw&\qw&\qw                           &\qw&\qw&\qw&\qw&\qw&\qw                           &\multigate{0}{>}&\ctrl{-8}&\ctrl{-12}&\qw&\qw&\qw &\qw\\
&\qw&\qw&\qw&\qw&\qw&\qw&\qw                           &\qw&\qw&\qw&\qw&\qw&\qw                           &\qw&\qw&\qw&\multigate{0}{>}&\ctrl{-7}&\ctrl{-11} &\qw
}
$$
\caption{ An example of a bitonic sort for $K=4$ and $6$ ancillas as introduced in \cite{antisym}. In each step, two registers encoding times are compared. If the compared values are in decreasing order, an ancilla qubit is flipped. The states of the involved registers are then swapped conditioned on the 
value of the ancilla qubit.  This results in sorting the values stored in a pair of registers. The same permutation is performed within the $\texttt{k}$ register. Note that by recording the result of each comparator in an ancilla qubit the sorting network is made reversible.} 
 \label{fig:bitonic}
\end{figure}

There are a number of possible sorting networks.
A simple example is the bitonic sort, and an example quantum circuit for that sort is shown in Fig.~\ref{fig:bitonic}.
Since we need to record the positions of the first $k$ registers as well, we perform the same controlled-swap on the $k$-register too.
The bitonic sort requires $\mathcal{O}(K \log^2 K)$ comparisons, but there are more advanced sorting networks that use $\mathcal{O}(K \log K)$ comparisons \cite{zig_zag}.
That brings the complexity of \texttt{Clock} preparation to $\mathcal{O}(K \log K \log M)$ elementary gates.
Since each comparison requires a single ancilla, the space overhead is $\mathcal{O}(K \log K)$ ancillas.

\subsection{Completing the State Preparation}
\label{sec:state-preparation-part3}
To complete the state preparation, we transform each of the $K$ registers encoding $\ell_1,\ldots\ell_K$ from $\ket{0}$ into equal superposition states.
If $L$ is a power of two, then this transformation can be taken to be just Hadamards on the individual qubits.
It is always possible to choose the value of $\gamma$ to make $L$ a power of two.
The complexity is then $\mathcal{O}(K\log L)$ single-qubit operations.
It is also possible to create an equal superposition state using $\mathcal{O}(K\log L)$ gates if $L$ is not a power of two, though the circuit is more complicated.

Next let us consider the state prepared as a result of this procedure.
In the first case, where we prepare the superposition of times by the compressed rotations, we first prepare the state $\ket{\Xi_M^K}$, which contains the separations between successive times, then add these to obtain the times.
The state is therefore
\begin{equation}
\sqrt{\frac{1-\mu^2}S} \sum_{k=0}^K \oldbeta^{k/2}\ket{k} \sum_{j_1<j_2\ldots<j_k} \ket{j_1,\ldots,j_k,M,\ldots,M}+\mu\ket{\nu''}\,.
\end{equation}
The preparation of the registers encoding $\ell_1,\ldots\ell_K$ gives a factor of $1/L^{K/2}$.
For $k<K$ the registers past $k$ are not used, and the effective amplitude factor is $1/L^{k/2}$.
We obtain the indices $j_1$ to $j_k$ in sorted order without repetitions.
This means that the weightings of the terms in the sum is $(\gamma T/(rM))^{k/2}$.
In Eq.~\eqref{betas}, when there are no repetitions the $k_1,\ldots,k_\sigma$ are all $1$.
Therefore we have approximately the desired state preparation from Eq.~\eqref{eq:state-preparation} with the correct weightings,
and with $s=S/(1-\mu^2)$.
There is imprecision of $\mathcal{O}(\epsilon/r)$ due to the additional term weighted by $\mu$, as well as imprecision due to the repetitions being omitted, which is bounded in Sec.~\ref{sec:ComplexityReq} below.

In order for amplitude amplification to take one step, we require that $s=S/(1-\mu^2)\le 2$.
Note that it can be less than $2$, and oblivious amplitude amplification can still be performed in a single step using an ancilla qubit, as noted in  the `Segment Lemma' of Ref.~\cite{BerrySTOC14}.
To bound the value of $S$,
\begin{align}
S&=\sum_{\left|x\right|\leq K} 
\oldbeta^{\left|x\right|} \nonumber \\
&= \sum_{k=0}^K
\binom{M}{k}\oldbeta^k \nonumber \\
&< \sum_{k=0}^\infty \frac{M^k\oldbeta^k}{k!} \nonumber \\
&= e^{M\oldbeta} = e^{\lambda T/r}.
\end{align}
Therefore, by choosing $r\ge\lambda T /\ln [2(1-\mu^2)]$ we can ensure that $s\le 2$, and a single step of amplitude amplification is sufficient.
The value of $\mu^2$ is $\mathcal{O}(\epsilon/r)$, and therefore $r=\Theta(\lambda T)$.

In the second case, where we obtain the superposition over times via a sort, the state is as in Eq.~\eqref{all_times} except sorted, with information about the permutation used for the sort in an ancilla.
When there are repeated indices $j$, the number of initial sets of indices that yield the same sorted set of indices is $k!/(k_1!k_2!\ldots k_\sigma!)$, using the same notation as in Eq.~\eqref{betas}.
That means for each sorted set of $j$ there is a superposition over this many states in the ancilla with equal amplitude, resulting in a multiplying factor of $\sqrt{k!/(k_1!k_2!\ldots k_\sigma!)}$ for each sorted set of $j$.

As noted above, because the times $t_{k+1}$ to $t_K$ are ignored, the factors of $M$ in the amplitude cancel.
Similarly, when we perform the state preparation for the registers encoding $\ell_1,\ldots\ell_K$, we obtain a factor of $1/L^{k/2}$.
Including these factors, and the factor for the superposition of states in the ancilla, we obtain amplitudes of
$(\gamma T/(rM))^{k/2}/\sqrt{k_1!k_2!\ldots k_\sigma!}$.
Hence we have the desired state preparation from Eq.~\eqref{eq:state-preparation} with the correct amplitudes, and the same $s$.

We require $s\le 2$ for oblivious amplitude amplification to take one step.
We can bound $s$ via
\begin{align}
s &= \sum_{k=0}^K \frac{M^k\oldbeta^k}{k!} \nonumber \\
&< \sum_{k=0}^\infty \frac{M^k\oldbeta^k}{k!} \nonumber \\
&= e^{M\oldbeta} = e^{\lambda T/r}.
\end{align}
Therefore, by choosing $r\ge \lambda T/ \ln 2$ we can ensure that $s\le 2$, and a single step of amplitude amplification is sufficient.
Note that it is convenient to take $r$ to be a power of two, so the qubits encoding the time may be separated into a set encoding the segment number and another set encoding the time within the segment.
Hence, in either case we choose $r=\Theta(\lambda T)$.

\section{Complexity requirements}\label{sec:ComplexityReq}
We now summarize the resource requirements of all the components of the algorithm and provide the overall complexity. The full quantum circuit consists of $r$ 
successively executed blocks, one for each of the 
$r$ time segments, respectively. The total cost is therefore $r$ times the cost of a single segment. 
Each segment requires 
one round of oblivious amplitude amplification, which includes two reflections $R$, two applications of $W$ and one application of the inverse $W^\dagger$, as in Eq.~\eqref{eq:OAA-transformation}. The cost of reflections is negligible compared to that of $W$. 
Hence, the overall cost of the algorithm amounts to 
$\mathcal{O}(r)$ times the cost of transformation $W$,  
whose quantum circuit is depicted in Fig.~\ref{fig:QCircuit-W}.
Implementing $W$ requires application of 
\texttt{clock} state preparation and its inverse, $2K$ applications of $\textsc{Had}^{\otimes \log L}$, and $K$ controlled applications of unitaries $H_\ell(t)$.

Choosing $K$ as in Eq.~\eqref{eq:scaling_KM} yields error due to truncation for each segment scaling as $\mathcal{O}(\epsilon/r)$, and therefore total error due to truncation scaling as $\mathcal{O}(\epsilon)$.
We choose $r=\Theta(\lambda T)$ for oblivious amplitude amplification, which means that $K$ is chosen as
\begin{equation}\label{Kchoice}
K =\Theta\left(\frac{\log(\lambda T/\epsilon)}{\log\log(\lambda T/\epsilon)}\right).
\end{equation}
Note that $\lambda\ge \maH$ implies that the condition $r\ge \maH T$ is satisfied.

It was stated above that the error due to discretization of the integrals is
$\mathcal{O} \left(  \frac{ (T/r)^2 \dot{H}_{\text{max}}}{M} \right)$.
That can be seen by noting that in time interval $T/(Mr)$ we have
\begin{align}
\| H(t) - H(t_j)\| &\le \frac{T}{rM}\max_z \left\|\frac{dH(z)}{dz}\right\| \nonumber \\
&\le \frac{T\mot}{rM}\,,
\end{align}
where $t_j$ is the time used in the discretization, and $\max_z$ indicates a maximum over that time interval.
Then the error in the integral over time $T/r$ is $\mathcal{O} \left(  \frac{ (T/r)^2 \dot{H}_{\text{max}}}{M} \right)$.
That is the error in the $k=1$ term for $\widetilde{U}$, and the error for terms with $k>1$ are higher order.
The error for all $r$ segments is then $\mathcal{O} \left(  \frac{ T^2 \dot{H}_{\text{max}}}{rM} \right)$.
We can ensure that the error due to the discretization of the integrals is $\mathcal{O}(\epsilon)$, by choosing
\begin{equation}\label{Mchoice1}
M\in \Theta\left( 
\frac{T \mot}{\epsilon \lambda } \right),
\end{equation}
where we have used $r\in \Theta(\lambda T)$.

In the case where the \texttt{clock} state is prepared using the compressed form of rotations, the operation that is performed is a little different than desired, because all repeated times are omitted.
For each $k$, the proportion of cases with repeated times is an example of the birthday problem, and is approximately $k(k-1)/(2M)$.
Therefore, denoting by $\widetilde{U}_{\rm unique}$ the operation corresponding to $\widetilde{U}$ but with repeated times omitted, we have
\begin{align}
\| \widetilde{U}- \widetilde{U}_{\rm unique}\| &\lesssim \sum_{k=2}^K \frac{(T/r)^k}{k!} \frac{k(k-1)}{2M}\maH^k \nonumber \\
&= \sum_{k=0}^K \frac{(T/r)^{k+2}}{k!} \frac{1}{2M}\maH^{k+2} \nonumber\\
&< \frac{T^2\maH^2}{2r^2M} e^{T\maH/(rM)}.
\end{align}
Therefore, over $r$ segments the error due to omitting repeated times is upper bounded by
\begin{equation}
r\| \widetilde{U}- \widetilde{U}_{\rm unique}\| \lesssim \frac{T^2\maH^2}{2rM} e^{T\maH/(rM)} \;.
\end{equation}
Using $r\in \Theta(\lambda T)$ and $\lambda \ge \maH$, we should therefore choose
\begin{equation}
M = \Omega \left( \frac{T\maH^2}{\epsilon\lambda } \right) .
\end{equation}
In the case where the repeated times are omitted, we would therefore take
\begin{equation}\label{Mchoice2}
M = \Theta \left( \frac{T}{\epsilon\lambda }(\maH^2+\dot{H}_{\text{max}}) \right) .
\end{equation}

Next we consider the complexity for the state preparation for the \texttt{clock} register.
In the case of the compressed rotation encoding, the complexity is $\mathcal{O}\left( K [\log{M}+\log\log(\lambda T/\epsilon)]\right)$, where we have used $r\in \Theta(\lambda T)$.
In typical cases we expect that the first term is dominant.
In the case where the \texttt{clock} register is prepared with a quantum sort, the complexity is $\mathcal{O}\left( K \log{M} \log{K}\right)$.

The preparation of the registers $\ell_1, \dots, \ell_k$ requires only creating an equal superposition. 
It is easiest if $L$ is a power of two, in which case it is just Hadamards.
It can also be achieved efficiently for more general $L$, and in either case the complexity is $\mathcal{O}\left( K\log{L}\right)$ elementary gates.

Next, one needs to implement the controlled unitaries $H_\ell$.
Each controlled $H_\ell$ can be implemented with $\mathcal{O}\left(1\right)$ queries to the oracles that give the matrix entries of the Hamiltonian \cite{BerrySTOC14}.
Since the unitaries are controlled by $\mathcal{O}\left( \log{L} \right)$ qubits and act on $n$ qubits, each control-$H_\ell$ requires $\mathcal{O}\left( \log{L} + n\right)$ gates.
Scaling with $M$ does not appear here, because the qubits encoding the times are just used as input to the oracles, and no direct operations are performed.
As there are $K$ controlled operations in a segment, the complexity of this step for a segment is
$\mathcal{O}\left(K\right)$ queries and $\mathcal{O}\left( K(\log{L} + n)\right)$ gates. 

In order to perform the simulation over the entire time $T$, we need to perform all $r$ segments, and since we take $r=\Theta(\lambda T)$ the overall complexity is multiplied by a factor of $\lambda T$.
For the number of queries to the oracle for the Hamiltonian, the complexity is
\begin{equation}
\mathcal{O}(\lambda T K)\,.
\end{equation}
Here $\lambda=L\gamma$ and $L$ is chosen as in Eq.~\eqref{Lsca} as $\Theta(d^2\maH/\gamma)$, so $\lambda\in\mathcal{O}(d^2\maH)$.
In addition, $K$ is given by Eq.~\eqref{Kchoice}, giving the query complexity
\begin{equation}
\mathcal{O}\left(d^2\maH T \frac{\log(d\maH T/\epsilon)}{\log\log(d\maH T/\epsilon)}\right).
\end{equation}

The complexity in terms of the additional gates is larger, and depends on the state preparation scheme used.
In the case where the preparation of the time registers is performed via the compressed encoding, we obtain complexity
\begin{equation}
\mathcal{O}\left( \lambda T K[\log{L}+\log{M}+\log\log(\lambda T/\epsilon) + n]\right).
\end{equation}
Regardless of the preparation, $\gamma$ is chosen as in Eq.~\eqref{gamchoice} as $\Theta(\epsilon/(d^3T))$.
That means $\lambda T/\epsilon = \Theta(L/d^3)$, so the double-logarithmic term $\log\log(\lambda T/\epsilon)$ can be omitted.
Moreover, $L\in\Theta(d^5\maH T/\epsilon)$, whereas the first term in the scaling for $M$ in Eq.~\eqref{Mchoice2} is $\Theta(\maH T/(\epsilon d^2))$.
This means that the first term in Eq.~\eqref{Mchoice2} can be ignored in the overall scaling due to the $\log L$, and we obtain overall scaling of the gate complexity as
\begin{equation}\label{gatecomplexity}
\mathcal{O}\left(d^2\maH T \frac{\log(d\maH T/\epsilon)}{\log\log(d\maH T/\epsilon)}\left[ \log\left(\frac{d \maH T}{\epsilon}\right) + \log\left( \frac{\mot T}{\epsilon d\maH}\right) + n \right]\right).
\end{equation}
There is also complexity of $\mathcal{O}(r\log r)$ for the increments of the register recording the segment number, but it is easily seen that this complexity is no larger than the other terms above.

In the case where the time registers are prepared using a sort, we obtain complexity
\begin{equation}
\mathcal{O}\left( \lambda T K(\log{L}+\log{M}\log{K} + n)\right),
\end{equation}
with $M$ given by Eq.~\eqref{Mchoice1}.
Technically this complexity is larger than that in Eq.~\eqref{gatecomplexity}, because of the multiplication by $\log K$.
Nevertheless, it is likely that in practice the preparation by a sort may be advantageous in some situations, because the value of $M$ that is required for the first preparation scheme is larger.

\section{Conclusion}
\label{sec:Conclusion}
We have provided a quantum algorithm for simulating the evolution generated by a time-dependent Hamiltonian that is 
given by a generic $d$-sparse matrix. 
The complexity of the algorithm scales logarithmically in both the dimension of the Hilbert space and the inverse of the error of approximation $\epsilon$.
This is an exponential improvement in error scaling compared to techniques based on Trotter-Suzuki expansion.
We utilize the truncated Dyson series, which is based on the truncated Taylor series approach by Berry \textit{et al.}~\cite{berry2015simulating}.
It achieves similar complexity, in that it has $T$ times a term logarithmic in $1/\epsilon$.
Interestingly, the complexity in terms of queries to the Hamiltonian is independent of the rate of change of the Hamiltonian.

For the complexity in terms of additional gates, the complexity is somewhat larger, and depends logarithmically on the rate of change of the Hamiltonian.
The complexity also depends on the scheme that is used to prepare the state to represent the times.
If one uses the scheme as in Ref.~\cite{berry2014gate}, which corresponds to a compressed form of a tensor product of small rotations, then there is additional error due to omission of repeated times.
Alternatively one could prepare a superposition of all times and sort them, which eliminates that error, but gives a multiplicative factor in the complexity.
This trade-off means that different approaches may be more efficient with different combinations of parameters.

One way in which this scheme is likely to be suboptimal is in the scaling with the square of the sparseness $d$.
That scaling comes from the decomposition of the Hamiltonian into 1-sparse Hamiltonians, and is worse than schemes based on quantum walks which are linear in $d$ \cite{BerryQIC12,BerryFOCS15}.
Another way that the complexity could potentially be improved is by the factor of $\log(1/\epsilon)$ being additive rather than multiplicative, as in Ref.~\cite{LowPRL2016}.
Those approaches do not appear directly applicable to the time-dependent case, however.

\acknowledgments
MK thanks Nathan Wiebe for insightful discussions. 
DWB is funded by an Australian Research Council Discovery Project (Grant No.\ DP160102426).

\bibliographystyle{hunsrt}
\bibliography{notes}

\end{document}